\documentclass{iopart}

\usepackage[dvips]{graphicx}

\newcommand{\erf}{\mathrm{erf}}
\newcommand{\erfc}{\mathrm{erfc}}

\newcommand{\sgn}{\mathrm{sgn}}

\newcommand{\D}{\rmd}

\newcommand{\Di}[1]{\!\!\rmd #1\,}

\newcommand{\E}{\ensuremath{\mathrm{e}}}

\newcommand{\Dee}{\ensuremath{\mathcal{D}}}

\newcommand{\avg}[1]{\ensuremath{\langle #1 \rangle}}
\newcommand{\avgU}[1]{\ensuremath{\langle #1 \rangle_U}}
\newcommand{\avgR}[1]{\ensuremath{\langle #1 \rangle_R}}

\newcommand{\psum}{\mathop{\sum\nolimits^\prime}}

\begin{document}

\title{Perturbation theory for the one-dimensional trapping reaction}
\author{R A Blythe and A J Bray} 
\address{Department of Physics and Astronomy, University of
Manchester, Manchester M13 9PL, U.K.}
\ead{richardb@theory.ph.man.ac.uk}

\address{21st August 2002}

\begin{abstract}
We consider the survival probability of a particle in the presence of
a finite number of diffusing traps in one dimension.  Since the
general solution for this quantity is not known when the number of
traps is greater than two, we devise a perturbation series expansion
in the diffusion constant of the particle.  We calculate the
persistence exponent associated with the particle's survival
probability to second order and find that it is characterised by the
asymmetry in the number of traps initially positioned on each side of
the particle.
\end{abstract}

\section{Introduction}

Our understanding of the nonequilibrium dynamics of many-body systems
has been greatly advanced through the study of particle reaction
models.  Typically these models involve a number of particles of
different species that undergo reactions which lead to a lack of
conservation of particle number.  Whilst the most obvious physical
application is to the kinetics of chemical reactions
\cite{Benson60,Rice85}, these models also enjoy mappings to a range of
phenomena.  These include interfacial growth \cite{KS91}, domain
coarsening in magnetic media and fluids \cite{Bray94,DZ96} and
aggregation \cite{Spouge88}.  Reaction-diffusion systems have also
held their own as prototypes for the development of theoretical tools,
such as the field theoretic renormalisation group \cite{Cardy97,MG98}
and exact methods in low dimensions \cite{Privman97}.

In this work we consider the dynamics of the $A+B\to\emptyset$
reaction, i.e.\ a system comprising two particle species that mutually
annihilate (or form an inert product) on contact.  This problem was
introduced in the seminal paper of Toussaint and Wilczek \cite{TW83}
as a model of monopole-antimonopole annihilation in the early
universe.  Since then, there have been applications to chemical
kinetics and condensed matter physics---see, for example,
\cite{Redner97} for a review.

In a reaction system, such as $A+B\to\emptyset$, one is chiefly
interested in the concentration of $A$ and $B$ particles after time
$t$ given some initial condition.  The `traditional' approach to the
problem is to write down differential \textit{rate equations} for the
density of each particle species, denoted $\rho_A(\vec{x},t)$ and
$\rho_B(\vec{x},t)$ respectively \cite{Benson60}.  One has for $\rho_A$
\begin{equation}
\label{rateeqn}
\frac{\partial \rho_A(\vec{x},t)}{\partial t} = 
D_A \nabla^2 \rho_A - r \rho_A \rho_B
\end{equation}
and the same equation for $\rho_B$ if one exchanges the labels $A$ and
$B$.  In words this equation says that both particle species diffuse
with diffusion constants $D_A$ and $D_B$ and particles are removed at
a rate proportional to the reaction constant $r$ and the joint
probability to find an $A$-$B$ pair at coordinate $\vec{x}$ at time
$t$.  Under the assumption that the particles are well-mixed at all
times, $\rho_{A,B}(\vec{x},t) = \rho_{A,B}(t)$ and one can neglect the
Laplacian in the rate equations.  In physics parlance, this is
equivalent to making a mean-field approximation which becomes exact in
the ``reaction-limited'' regime where the reaction rate is slow
compared to the diffusion rate.  Here we are interested in the
opposite ``diffusion-limited'' regime.

The long-time form of $\rho_A$ and $\rho_B$ depends on whether the
initial numbers of $A$ and $B$ particles are equal or not.  Consider
first the case $\rho_A(0)=\rho_B(0)$.  Since each reaction event
removes both an $A$ and a $B$ particle, their densities remain equal
at all times.  Then, we have for $\rho(t) = \rho_A(t) = \rho_B(t)$
\begin{equation}
\frac{\D \rho(t)}{\D t} = - r \rho^2
\end{equation}
whose solution at large times is $\rho(t) \sim 1/(r t)$.  It turns out
that this behaviour holds only in more than four dimensions.  Through
rigorous bounding arguments \cite{BL88} and renormalisation group
analysis \cite{LC95} it is now well understood that in dimensions
$d<4$ the density decays as $\rho(t) \sim 1/t^{d/4}$.  These results
serve as a reminder of the importance of density fluctuations and
correlations in low dimensions, which were explicitly ignored in the
mean-field approach.  Also we note that, in this problem, there exists
a \textit{dynamical scaling regime} in which the characteristic
length-scale in the problem (here the interparticle separation $\ell =
\rho^{-1/d}$) is related to a characteristic time-scale through a
power-law.

When the initial densities of $A$ and $B$ particles are not equal, the
situation is somewhat different.  Consider an initial condition that
has $\rho_A(0) < \rho_B(0)$.  Then, as time progresses the
$A+B\to\emptyset$ recation implies that the ratio
$\rho_A(t)/\rho_B(t)$ decreases.  Eventually one has a small number of
$A$ particles in a sea of $B$ particles whose density remains
effectively constant.  In the rate-equation formalism (\ref{rateeqn})
we have, under the assumption that particles are well-mixed and
$\rho_B(t) = \mbox{const} = \rho_B$, that the density of $A$ particles
decays as $\rho_A(t) \sim \exp( - r \rho_B t)$.  Again it is known that
this result holds only in a space of suitably large dimensionality.
In particular Bramson and Lebowitz \cite{BL88} used bounding arguments
to show that, at long times, the $A$ particles experience a
stretched-exponential decay $\rho_A(t) \sim \exp(-\lambda_d t^{d/2})$
in fewer than two dimensions, and $\rho_A(t) \sim\exp(-\lambda_2 t/\ln
t)$ in the critical dimension $d=2$.  Values of the constants
$\lambda_d$ for $d \le 2$ were recently reported by us \cite{BB02a}
with the details to be presented elsewhere \cite{BB02c}.

Clearly the special case of a single $A$ particle in the presence of a
sea of $B$ particles will be governed by these asymptotics.  Then, one
can interpret the $\rho_A(t)$ as the \textit{survival probability} of
the $A$ particle.  If the distribution of $B$ particles is homogeneous
over all space, and the diffusion constants of the $A$ and $B$
particles are the same, one can also view $\rho_A(t)$ as the
fraction of particles that have not met any other particles.  Thus the
reaction $A+B\to \emptyset$ in the limit of a low density of $A$
particles has been discussed under the guises of \textit{uninfected
walkers} \cite{OB02} in which random walkers infect each other on
contact, \textit{diffusion in the presence of traps} \cite{DV75,MG02} in
which the $B$ particles are considered as traps for the $A$ particles,
and \textit{predator-prey models} \cite{RK99} in which one asks for
the survival of a prey (the $A$ particle) being `chased' by diffusing
predators (the $B$ particles).  To avoid confusion, we shall use only
the trapping terminology, and in this work we restrict ourselves to
the case of a single $A$ particle and a \textit{finite} number of
diffusing traps ($B$ particles) in one spatial dimension.  When we
make no distinction between the particle and the traps, we shall use
the generic term \textit{walkers}.

Before we formally introduce our model and review some exactly known
results, let us make a few comments about the relationship between the
trapping reaction and \textit{persistence} in diffusive systems.
Persistence is a property that can be defined for any stochastic
dynamical system as it is simply the probability that a random
variable does not change sign between time $0$ and a time $t$.  In
many cases, a persistence probability $Q(t)$ decreases as a power-law
in time, $Q(t) \sim t^{-\theta}$ in which $\theta$ is a
\textit{persistence exponent} (see, e.g., \cite{Majumdar99} for a
review).  In the trapping reaction we can define two distinct
persistence properties: (i) a \textit{site} persistence, which
measures the probability that a particular point in space (or site on
a lattice) has not been crossed by any walker; and (ii) a
\textit{walker} persistence, which is the probability that a
particular walker (here the $A$ particle) has not yet met another
walker.  In this terminology, one can make the transition from walker
to site persistence by decreasing the particle's diffusion constant to
zero.  Then the particle is static, and the probability that it
survives is equal to the probability that a particular point has not
been crossed.

In the following section we will show that the site persistence
probability in the presence of $N$ traps in one dimension is known
exactly to decay with an exponent $\theta = N/2$.  On the other hand
the more general walker persistence probability, where the particle
has a nonzero diffusion constant, is not known exactly.  The principal
purpose of this paper is present the full details of a perturbation
theory for the walker persistent exponent that was outlined in a
previous work \cite{BB02a}.  As well as describing the approach in
greater detail, we also extend the calculation to second order in (a
quantity closely related to) the particle's diffusion constant.  We
find that the persistence exponent depends on a quantity that
characterises the level of asymmetry between the number of traps
initially positioned to each side of the particle.

To close this introductory section, we make a few remarks as to why
the case where both the particle and the traps diffuse is considerably
more complicated than that in which either species is static (the case
of static traps is sometimes referred to as the Donsker-Varadhan
problem \cite{DV75} for which the walker persistence decays as a
stretched exponential).  When the walker is static, the probability
that none of the $N$ independently diffusing traps has crossed a
particular point is simply the product of probabilities for each of
the traps separately not to have crossed that point.  However, any
movement of the particle (in the subset of trajectories for which the
particle survives) correlates the motion of the traps, and so the
probabilities for each of the traps not to have met the particle are
no longer independent.  To understand this point more clearly,
consider a set of walkers on the one-dimensional lattice.  When the
particle takes a step to the left (say), all the traps to the left hop
one site closer to the particle in concert, whereas those to the right
all hop one site away.  It is this collective motion of the traps in
the particle's frame of reference that makes the problem very
difficult mathematically, and as far as we are aware, exact
expressions for the walker persistence exponent are known only when $N
\le 2$.

In the following section, we discuss these known results more
quantitatively, and make the connection to isotropic diffusion in an
$N$-dimensional `hyperwedge' geometry.  Then, in
section~\ref{secapproach}, we describe how one sets up a perturbation
theory using a path-integral approach.  A judicious choice of
timescale reveals that the natural parameter in the perturbation
expansion has a geometric significance in the hyperwedge problem.  In
section~\ref{seccalc} we go on to calculate the coefficients of the
first and second order terms in the perturbation expansion, before
concluding with some discussion and open questions.

\section{Model definition and review of known results}
\label{secdef}

The model we consider is defined as follows.  At time $t=0$, a
particle is placed at the origin of a one-dimensional space
$x_0(0)=0$.  Additionally, a set of mobile traps are placed at
positions $x_i(0)=y_i$ with $i=1,2,\ldots, N$ such that $N$ is the
number of traps.  We will at times need to distinguish between those
traps initially placed to the left and right of the particle.  To this
end we introduce the shorthand $\sigma_i = \sgn(y_i)$ and the
quantities $N_L$ and $N_R$ which are the number of particles that have
$\sigma_i<0$ and $\sigma_i>0$ respectively.

The system evolves through the independent diffusion of the particle
and traps.  We shall take the diffusion constant of the particle to be
$D^\prime$ and that of the traps to be $D$.  Then, we can write a
Langevin equation
\begin{equation}
\label{langevin}
\dot{x}_i = \eta_i(t)
\end{equation}
in which $\eta_i(t)$ is a Gaussian white noise with zero mean and
correlator
\begin{equation}
\label{correlator}
\avg{\eta_i(t) \eta_j(t^\prime)} = 2 D_i \delta_{i,j}
\delta(t^\prime-t) \;,
\end{equation}
where $D_0 = D^\prime$ and $D_i = D$ for $i \ne 0$.

If, at any time, a trap coordinate $x_i(t)$ coincides with the
particle coordinate $x_0(t)$, the particle becomes `trapped'.  We seek
an expression for the \textit{survival probability} $Q(t)$, i.e.\ the
probability that the particle has not yet fallen into a trap.

It is convenient to work in the particle's frame of reference and so
we introduce the relative coordinates $X_i=x_i-x_0$ and $Y_i=y_i$.
Then the connection to persistence becomes clearer, since $Q(t)$ is
the probability that none of the variables $X_i$ has changed sign
until a time $t$.  From (\ref{langevin}) and (\ref{correlator}) we
obtain the underlying Langevin equations in the variables $X_i$.
These read
\begin{equation}
\label{langevin2}
\dot{X}_i = \xi_i(t)
\end{equation}
in which the noise $\xi_i(t)$ has mean zero and correlator
\begin{equation}
\label{xicorrelator}
\avg{\xi_i(t) \xi_j(t^\prime)} = 2 (D \delta_{i,j} +
D^\prime) \delta(t^\prime-t) \equiv 2 D_{ij} \delta(t^\prime-t) \;.
\end{equation}
Here we have introduced the matrix $D_{ij} = D \delta_{i,j} +
D^\prime$ which expresses the way in which the trap's trajectories
are correlated in the frame of reference of a diffusing particle
(i.e.\ when $D^\prime \ne 0$).

In the introduction, we remarked that the survival probability $Q(t)$
can be calculated exactly when $D^\prime=0$.  Although this is a
classic result (see, e.g., \cite{Weiss94}), we present the
details of the calculation here as it is will be needed when
calculating the coefficients in the perturbation expansion that we
construct for general $D^\prime$ in the next section.

We begin by writing down the Fokker-Planck equation for the
probability distribution function $G_N( \{ X \}, t| \{Y \})$ that
holds when $D^\prime = 0$.  It reads
\begin{equation}
\label{FP-N}
\frac{\partial}{\partial t} G_N( \{ X \}, t| \{Y \}) =
D \sum_{i=1}^N \frac{\partial^2}{\partial X_i^2} G_N ( \{ X \}, t|
\{Y\})
\end{equation}
and is subject to the absorbing boundary condition $G_N = 0$ if any of
the $X_i$ are zero (i.e., when a trap meets the stationary particle).
Clearly, since the traps diffuse independently, the solution of this
equation is just
\begin{equation}
\label{Gprod}
G_N(\{X\},t|\{Y\}) = \prod_{i=1}^{N} G_1(X_i,t| Y_i)
\end{equation}
where $G_1(X,t; Y)$ is the probability for a walker starting at
$Y$ to be at $X$ and not to have crossed the origin up to time $t$.
For a particle starting at $Y>0$, this quantity is given by
\begin{equation}
\label{G1}
G_1(X,t| Y) = \frac{1}{\sqrt{4\pi D t}} \left[ \exp \left(
-\frac{(X-Y)^2}{4Dt} \right) - \exp \left( -\frac{(X+Y)^2}{4Dt}
\right) \right] 
\end{equation}
for $X>0$.  We note that this is the correct solution since the two
terms on the right-hand side separately obey (\ref{FP-N}) with $N=1$
and this particular combination of solutions satisfies the boundary
condition $G_1(0,t| Y)=0$.

To obtain the probability for a walker starting at $Y$ not to have
crossed the origin until time $t$ we simply integrate (\ref{G1}) over
all $X>0$ to find
\begin{equation}
Q_1(t | Y) = \erf\left( \frac{Y}{2\sqrt{Dt}} \right) \;.
\end{equation}
Note that, by symmetry, we must have $Q_1(t | -Y) = Q_1(t|Y)$ so the
general result is
\begin{equation}
\label{Q1}
Q_1(t | Y) = \erf\left( \frac{|Y|}{2\sqrt{Dt}} \right) \;.
\end{equation}
Finally, using (\ref{Gprod}) we find the probability that none of the
traps has met the particle is
\begin{equation}
\label{QN}
Q_N(t) = \prod_{i=1}^N \erf\left( \frac{|Y_i|}{2\sqrt{Dt}} \right)
\sim \left(\frac{1}{\pi D t}\right)^{N/2} \prod_{i=1}^N |Y_i|
\end{equation}
in which we have used the fact that
\begin{equation}
\erf x = \frac{2}{\sqrt{\pi}} \left( z - \frac{z^3}{3} + \cdots \right)
\end{equation}
to obtain a form for $Q_N(t)$ valid at large times.  The persistence
exponent $\theta$ associated with the survival probability $Q(t)$ can
be defined formally as
\begin{equation}
\label{thetadef}
\theta = - \lim_{t\to\infty} t \frac{\D}{\D t} \ln Q(t)
\end{equation}
and so for the case $D^\prime=0$ we have from (\ref{QN}) that $\theta
= N/2$.

It is instructive to consider an approach that has been used (see
\cite{RK99,Redner01} for a more in-depth overview) to find the
persistence exponent $\theta$ exactly for the case $N=2$ and $D^\prime
\ne 0$ since it will reveal a connection with the perturbation series
that we derive in the next section.

The essence of this approach is to perform a second coordinate
transformation that renders the correlated diffusion described by
equations (\ref{langevin2}) and (\ref{xicorrelator}) isotropic in the
$N$-dimensional space spanned by the coordinates $X_i$.  To perform
this transformation one considers the diffusion matrix $D_{ij} = D
\delta_{i,j} + D^\prime$.  The eigenvectors of this matrix indicate
the directions in which the $N$-dimensional diffusion is independent
whilst the corresponding eigenvalues give the diffusion constants in
those directions.  One finds that one of the eigenvectors is
$(1,1,1,\ldots,1)$ and has an eigenvalue $D+N D^\prime$.  The
remaining eigenvectors are degenerate with eigenvalue $D$ and hence
form a basis set in the $(N-1)$-dimensional space perpendicular to
$(1,1,1,\ldots,1)$.  In order to make the diffusion isotropic, one
must scale in the $(1,1,1,\ldots,1)$ direction by a factor
$(1+ND^\prime/D)^{-1/2}$.

Whilst this coordinate transformation gives rise to simple, isotropic
diffusion in the $N$-dimensional space, the boundary conditions become
more complicated.  In the space spanned by the variables $X_i$, the
absorbing boundary is constructed from the set of orthogonal planes
that have $X_i=0$ with $i=1,2,\ldots, N$.  The coordinate transformation
just described rotates these planes so that they are no longer
orthogonal.  It is a straightforward exercise in linear algebra to
determine that the angle between two planes $i$ and $j$ in the
transformed coordinate system is $\Theta_{ij} = \arccos ( -\sigma_i
\sigma_j \alpha_N )$ in which the parameter $\alpha_N$ takes the form
\begin{equation}
\label{alphadef}
\alpha_N = \frac{D^\prime}{D + (N-1) D^\prime} \;.
\end{equation}
Hence two planes that correspond to traps that start on opposite sides
of the particle close up whereas two planes that correspond to traps
starting on the same side of the particle open out.

To determine the persistence exponent $\theta$ given by equation
(\ref{thetadef}) it is necessary to solve the diffusion equation in an
$N$-dimensional wedge geometry.  As far as we know, results are
available only for the case of a two-dimensional wedge
\cite{FG88,Redner01}.  Then, the particle's survival probability
decays with the exponent
\begin{equation}
\theta = \frac{\pi}{2 \Theta_{12}} \;.
\end{equation}
For the case where both traps initially surround the particle, we 
have $\sigma_1 \sigma_2 = -1$ and so
\begin{equation}
\label{exactsurround}
\theta = \frac{\pi}{2 \arccos \alpha_2} \;.
\end{equation}
Similarly when both traps are initially positioned to one side of the
particle, we have $\sigma_1 \sigma_2 = 1$ and thus
\begin{equation}
\label{exactchase}
\theta = \frac{\pi}{2 ( \pi - \arccos \alpha_2 )}
\end{equation}
since $\arccos(-x) = \pi - \arccos(x)$.

Although this approach has thus far proved intractable for $N>2$, we
have spent some time outlining it because, in the perturbation
theory that we construct below, it turns out that the natural
expansion parameter is the quantity $\alpha_N$ introduced above.  Note
that $\alpha_N=0$ corresponds to the case where the particle is static
and the $N$-dimensional hyperwedge is in fact a corner of an
$N$-dimensional hypercube.  Thus the expansion for small $\alpha_N$
derived below using an alternative path-integral approach gives results
for diffusion in a geometry perturbed from an $N$-dimensional
hypercube.

\section{Perturbation expansion via a path-integral approach}
\label{secapproach}

The starting point in the derivation of a perturbation expansion for
the persistence exponent $\theta$ is to note that the statistical
weight a particular set of trajectories $\vec{X}(t) = [X_1(t), X_2(t),
\ldots, X_N(t)]$ governed by equations (\ref{langevin2}) and
(\ref{xicorrelator}) can be written in the form $\exp(-S[\vec{X}])$.
In this expression, $S[\vec{X}]$ is an `action' functional defined as
\begin{equation}
S[\vec{X}] = \frac{1}{4} \sum_{i,j=1}^{N} [D^{-1}]_{ij} \int_0^t
\Di{t^\prime} \dot{X}_i(t^\prime) \dot{X}_j(t^\prime)
\end{equation} 
in which $D^{-1}$ is the inverse of the matrix $D$ in equation
(\ref{xicorrelator}).  It is straightforward to show that
\begin{equation}
[D^{-1}]_{ij} = \frac{1}{D} \left( \delta_{ij} - \frac{D^\prime}{D + N
D^\prime} \right)\;.
\end{equation}
It is helpful at this stage to introduce a rescaled time variable
\begin{equation}
\tau = \frac{D+ND^\prime}{D+(N-1)D^\prime} \, Dt
\end{equation}
such that the action becomes
\begin{equation}
S[\vec{X}] = S_0 - \frac{D^\prime}{D+(N-1)D^\prime} S_1 = S_0 -
\alpha_N S_1
\end{equation}
in which the diagonal part $S_0$ and off-diagonal part $S_1$ of the
action are given by
\begin{eqnarray}
S_0[\vec{X}] &=& \frac{1}{4} \sum_{i} \int_0^{\tau} \Di{\tau^\prime}
[\dot{X}_i(\tau^\prime)]^2 \\
S_1[\vec{X}] &=& \frac{1}{4} \sum_{i\ne j} \int_0^{\tau} \Di{\tau^\prime}
\dot{X}_i(\tau^\prime) \dot{X}_j(\tau^\prime)
\end{eqnarray}
where now a dot denotes differentiation to the rescaled time variable
$\tau$.  Note that when $\alpha_N=0$ the action
$S[\vec{X}]=S_0[\vec{X}]$ which gives the statistical weight for $N$
walkers executing \textit{independent} diffusive motion with diffusion
constant unity.  We will treat the off-diagonal part (that describes
the correlations between the walkers) as a perturbation with the
quantity $\alpha_N$ given by equation~(\ref{alphadef}) as the
expansion parameter, as advertised.

To this end, observe that out of all possible trajectories
$\vec{X}(\tau)$, only a subset corresponds to the particle surviving
until time $\tau$.  These trajectories are defined through the
persistence condition that $\sgn X(\tau^\prime) = \sgn X(0)$ for $0 <
\tau^\prime \le \tau$.  We shall use the symbol $\int_R
\Dee{\vec{X}(\tau)}$ to denote integration over this restricted set of
surviving trajectories.  Then, the probability that the particle
survives is
\begin{equation}
\label{path1}
Q(\tau) = \frac{\int_R \Dee{\vec{X}(\tau)} \exp(-S[\vec{X}])}%
{\int \Dee{\vec{X}(\tau)} \exp(-S[\vec{X}])}
\end{equation}
in which the integral of the weights over the restricted trajectories
is normalised by the integral over all possible trajectories.

Now, note that the survival probability $Q_N(\tau)$ of a stationary
particle in the presence of $N$ traps diffusing independently with
diffusion constant unity is given by the path-integral expression
\begin{equation}
Q_N(\tau) = \frac{\int_R \Dee{\vec{X}(\tau)} \exp(-S_0[\vec{X}])}%
{\int \Dee{\vec{X}(\tau)} \exp(-S_0[\vec{X}])} \;.
\end{equation}
Combining this with (\ref{path1}) we find that
\begin{eqnarray}
\label{Qtau}
\fl Q(\tau) = Q_N(\tau) \, \frac{\int_R \Dee{\vec{X}(\tau)}
\exp(-S[\vec{X}])}{\int_R \Dee{\vec{X}(\tau)} \exp(-S_0[\vec{X}])} \,
\frac{\int \Dee{\vec{X}(\tau)} \exp(-S_0[\vec{X}])}{\int \Dee{\vec{X}(\tau)}
\exp(-S[\vec{X}])}\nonumber\\
\lo= Q_N(\tau) \frac{\avgR{\rme^{\alpha_N
S_1[\vec{X}]}}}{\avgU{\rme^{\alpha_N S_1[\vec{X}]}}}
\end{eqnarray}
in which the notation $\avg{\cdot}$ indicates an average
over paths weighted by $\exp(-S_0[\vec{X}])$ and the subscripts $R$
and $U$ denote the restricted and unrestricted ensembles of paths.
(Recall that paths that cross the origin are excluded from the
former).

To obtain a perturbative expression for the persistence exponent
$\theta$ we make use of (\ref{thetadef}) and (\ref{Qtau}) to find, in
the rescaled time variable $\tau$,
\begin{eqnarray}
\fl \theta = - \lim_{\tau \to \infty} \tau \frac{\D}{\D \tau} \ln Q(\tau)
\nonumber\\
\lo= - \lim_{\tau \to \infty} \tau \frac{\D}{\D \tau} \left( \ln
Q_N(\tau) + \ln \avgR{ \rme^{\alpha_N S_1[\vec{X}]} } - \ln \avgU{
\rme^{\alpha_N S_1[\vec{X}]}} \right) \;.
\end{eqnarray}
The asymptotic form of $Q_N(\tau)$ is known from equation (\ref{QN}).
To compute the averages of $S_1[\vec{X}]$ perturbatively, we make use
of the fact that $\ln \avg{\exp \lambda x}$ defines the generating
function of the cumulants of $x$.  Specifically
\begin{equation}
\label{cumexp}
\ln \avg{\rme^{\lambda x}} = \lambda \avg{x} + \frac{\lambda^2}{2}
\left( \avg{x^2} - \avg{x}^2 \right) + \mathrm{O}
\left(\lambda^3\right) \;.
\end{equation}
Putting all this together, we find that $\theta$ has a series
expansion
\begin{equation}
\label{thetaexpand}
\fl \theta = \frac{N}{2} - \lim_{\tau \to \infty} \tau \frac{\D}{\D \tau}
\left[ \alpha_N \left( \avgR{S_1} - \avgU{S_1} \right) +
\frac{\alpha_N^2}{2} \left( \avgR{S_1^2}- \avgR{S_1}^2 - \avgU{S_1^2}
+ \avgU{S_1}^2 \right) \right]
\end{equation}
to second order in the parameter $\alpha_N$ defined in equation
(\ref{alphadef}).  Of course, one could go to higher order by
including more terms from the cumulant expansion (\ref{cumexp}).

\section{Walker persistence exponent to second order}
\label{seccalc}

As a first step in computing the mean and variance of $S_1$ in the two
ensembles we shall determine the structure of these quantities.  For
either the restricted or the unrestricted average we have
\begin{equation}
\avg{S_1} = \frac{1}{4} \psum_{i,j} \int_0^\tau
\Di{\tau^\prime} \avg{\dot{X}_i(\tau^\prime) \dot{X}_j(\tau^\prime)}
\end{equation}
in which the notation $\psum$ means the sum over combinations of the
indices such that each index is different.  (Note that all
permutations of a set of distinct indices are included in this sum).
Since we are averaging over ensembles of trajectories of
\textit{independently} diffusing particles, and the indices $i$ and
$j$ are always different, we have
\begin{equation}
\label{mean}
\avg{S_1} = \frac{1}{4} \psum_{i,j} \int_0^\tau \Di{\tau^\prime}
\avg{\dot{X}_i(\tau^\prime)} \avg{\dot{X}_j(\tau^\prime)} \;.
\end{equation}
Hence to calculate the mean of $S_1$ we must find the mean velocity of
a random walk with diffusion constant unity in each of the ensembles.

For the variance of $S_1$ we have
\begin{eqnarray}
\fl \avg{S_1^2} - \avg{S_1}^2 = \frac{1}{16} \psum_{i,j} \psum_{k,\ell}
\int_0^\tau \Di{\tau_1} \int_0^\tau \Di{\tau_2} \left( \avg{\dot{X}_i(\tau_1)
\dot{X}_j(\tau_1) \dot{X}_k(\tau_2) \dot{X}_\ell(\tau_2)} \right.
\nonumber\\
- \left. \avg{\dot{X}_i(\tau_1) \dot{X}_j(\tau_1)}\avg{\dot{X}_k(\tau_2)
\dot{X}_\ell(\tau_2)} \right) \;.
\end{eqnarray}
The terms in the double summation can be divided into three groups.
In the first, all four indices are different, and two averages
factorise to cancel.  A second set of terms has two indices the same,
the other two different (there are four ways this property can be
satisfied).  Finally there are two ways to arrange for the indices to
comprise two pairs the same.  Thus the variance can be written as
\begin{eqnarray}
\label{variance}
\fl \avg{S_1^2} - \avg{S_1}^2 = 
\frac{1}{16} \int_0^\tau \Di{\tau_1} \int_0^\tau \Di{\tau_2}
\nonumber\\
\fl\quad \left[ 4 \psum_{i,j,k} \left( \avg{\dot{X}_i(\tau_1)
\dot{X}_i(\tau_2)} - \avg{\dot{X}_i(\tau_1)}\avg{\dot{X}_i(\tau_2)}
\right) \avg{\dot{X}_j(\tau_1)} \avg{\dot{X}_k(\tau_2)}
\right. \nonumber\\
\fl\quad \left. + 2 \psum_{i,j} \left( \avg{\dot{X}_i(\tau_1)
\dot{X}_i(\tau_2)} \avg{\dot{X}_j(\tau_1) \dot{X}_j(\tau_2)} -
\avg{\dot{X}_i(\tau_1)} \avg{\dot{X}_i(\tau_2)}
\avg{\dot{X}_j(\tau_1)}\avg{\dot{X}_j(\tau_2)} \right) \right]
\end{eqnarray}
which reveals that we must calculate not just the mean velocity
$\avg{\dot{X}_i(\tau^\prime)}$ but also the velocity correlation function
$\avg{\dot{X}_i(\tau_1) \dot{X}_i(\tau_2)}$ over the two ensembles.

The unrestricted averages are easily found.  Since each of the walkers
performs diffusion with a diffusion constant of unity in the rescaled
time $\tau$, we have $\dot{X}_i(\tau^\prime) = \eta_i(\tau^\prime)$
where $\eta_i(\tau^\prime)$ is a Gaussian white noise with zero mean
and a correlator $\avg{ \eta_i(\tau_1) \eta_i(\tau_2) } = 2
\delta(\tau_2-\tau_1)$.  Hence
\begin{equation}
\label{meanvarU}
\avgU{\dot{X}_i(\tau^\prime)} = 0 \quad \mbox{and} \quad
\avgU{\dot{X}_i(\tau_1) \dot{X}_i(\tau_2)} = 2 \delta(\tau_2-\tau_1) \;.
\end{equation}

The restricted averages involve more work, since we must calculate the
mean velocity of a walker at time $\tau^\prime$ \textit{given} that it
survives until a later time $\tau$; similarly we must include the fact
that the walker survives until a specified time $\tau$ in the
calculation of the velocity correlation function.  Before getting into
the details, therefore, let us propose their general form and the
implications of such a form in context of the expression
(\ref{thetaexpand}).  On dimensional grounds, we suggest that
\begin{eqnarray}
\label{guess1}
\avgR{\dot{X}_i(\tau^\prime)} = \frac{1}{\sqrt{\tau^\prime}}
F_i(\tau_0/\tau^\prime, \tau^\prime/t)\\
\label{guess2}
\avgR{\dot{X}_i(\tau_1) \dot{X}_i(\tau_2)} = 2 \left[ \delta(\tau_2-\tau_1)
+ \frac{1}{\tau_+} G_i( \tau_0/\tau_-, \tau_-/\tau_+, \tau_+/\tau ) \right]
\end{eqnarray}
in which $\tau_0$ is some early timescale in the problem and $\tau_-$
($\tau_+$) are the smaller (respectively larger) of $\tau_1$ and $\tau_2$.
Guided by (\ref{meanvarU}), and confirmed by explicit calculation to
be presented below, we have included a delta-function contribution in
the two-time correlation function.

The upshot of this is that all the integrals in (\ref{mean}) and
(\ref{variance}) reduce to one of two forms.  These are
\begin{eqnarray}
\int_0^\tau \frac{\D{\tau^\prime}}{\tau^\prime} f(\tau_0/\tau^\prime,
\tau^\prime/\tau) \\
\int_0^\tau \frac{\D{\tau_+}}{\tau_+^2} \int_0^{\tau_+} \Di{\tau_-}
g(\tau_0/\tau_-, \tau_-/\tau_+, \tau_+/\tau)
\end{eqnarray}
in which the functions $f$ and $g$ are various combinations of the
velocity and velocity-correlation functions.  It is not the values of
these integrals that we are interested in, but the limit that appears
in the perturbation expansion (\ref{thetaexpand}).  It is
straightforward to show that
\begin{eqnarray}
\label{limit1}
\fl \lim_{\tau\to\infty} \tau \frac{\D}{\D{\tau}} \int_0^\tau
\frac{\D{\tau^\prime}}{\tau^\prime} f(\tau_0/\tau^\prime,
\tau^\prime/\tau) = f(0,0)\\
\label{limit2}
\fl \lim_{\tau\to\infty} \tau \frac{\D}{\D{\tau}} \int_0^\tau
\frac{\D{\tau_+}}{\tau_+^2} \int_0^{\tau_+} \Di{\tau_-}
g(\tau_0/\tau_-, \tau_-/\tau_+, \tau_+/\tau) = \int_{0}^{1} \Di{u}
g(0,u,0) \;.
\end{eqnarray}
The fact that zeros appear in the arguments of the functions $f$ and
$g$ implies that when calculating $\avgR{\dot{X}_i(\tau^\prime)}$ and
$\avgR{\dot{X}_i(\tau_1) \dot{X}_i(\tau_2)}$ we need only determine
their forms in the regime $\tau_0 \ll \tau^\prime, \tau_1, \tau_2 \ll
\tau$ to obtain the expansion (\ref{thetaexpand}).

We now indicate how to calculate these quantities.  First, note that
$\avgR{\dot{X}_i(\tau^\prime)} = \frac{\D}{\D \tau^\prime}
\avgR{X_i(\tau^\prime)}$.  This reduces the problem to the calculation of
the mean position of a walker at time $\tau^\prime$ taking into account
that it does not cross the origin (survives) until at least a time
$\tau$.  For the case where a walker's initial position $Y_i > 0$, this
quantity can be expressed as
\begin{equation}
\label{Xavg}
\avgR{X_i(\tau^\prime)} = \int_0^\infty \Di{X_i} X_i \frac{P_1(\tau;
X_i, \tau^\prime | Y_i)}{Q_1(\tau | Y_i)}
\end{equation}
in which $P_1(\tau; X_i, \tau^\prime | Y_i)$ is the probability that a
single walker visits $X_i$ at time $\tau^\prime$ and does not visit
the origin at any time $0 \le \tau^\prime \le \tau$.  In order to
average $X_i$ with respect to this distribution, it has been
normalised through the factor $Q_1(\tau | Y_i)$, which is the
probability the walker survives until time $\tau$ and is given by
equation (\ref{Q1}).

Since the diffusion process is Markovian, we can perform the
factorisation $P_1(\tau; X_1, \tau^\prime | Y_i) =
Q_1(\tau-\tau^\prime | X_i) G_1(X_i, \tau^\prime | Y_i)$, where we
recall that $G_1(X_i, \tau^\prime | Y_i)$ is that probability that a
walker is at $X_i$ at time $\tau^\prime$ given that it started at
$Y_i$ and whose form is given by equation (\ref{G1}).  Inserting the
explicit expressions for $Q_1(\tau| Y)$ and $G_1(X, \tau | Y)$ into
(\ref{Xavg}) we find
\begin{equation}
%
%
\fl \avgR{ X_i(\tau^\prime) } = \int_0^\infty \Di{X_i} X_i \frac{\erf \Big(
\frac{X_i}{2\sqrt{\tau-\tau^\prime}} \Big) \bigg[ \exp \Big(
-\frac{(X_i-Y_i)^2}{4 \tau^\prime} \Big) - \exp \Big(
-\frac{(X_i+Y_i)^2}{4 \tau^\prime} \Big) \bigg]}%
{\sqrt{4\pi  \tau^\prime} \erf \Big( \frac{Y_i}{2\sqrt{\tau}}
\Big)} \;.
\end{equation}

Now we can use the fact that when calculating the exponent $\theta$
using (\ref{thetaexpand}) it is sufficient to know
$\avgR{X_i(\tau^\prime)}$ in the regime $\tau_0 \ll \tau^\prime \ll
\tau$.  We take $\tau_0 = Y_i^2$ (recalling that the rescaled time
variable has units of length squared) and on expanding the integrand
in the previous expression, we find
\begin{equation}
\avgR{X_i(\tau^\prime)} = \frac{1}{2 \sqrt{\pi} ( \tau^\prime)^{3/2}}
\int_0^\infty \Di{X_i} X_i^3 \exp\left( - \frac{X_i^2}{4 \tau^\prime}
\right) = 4 \sqrt{\frac{\tau^\prime}{\pi}} 
\end{equation}
in this regime.  Finally, we differentiate to obtain
\begin{equation}
\label{meanR}
\avgR{\dot{X}_i(\tau^\prime)} = \frac{2}{\sqrt{\pi \tau^\prime}}
\end{equation}
when $\tau_0 \ll \tau^\prime \ll \tau$ and $Y_i>0$.  Since we are
working at times much larger than $\tau_0 = Y_i^2$, the dependence on
the magnitude of $Y_i$ has dropped out of this expression.  However,
the sign of the average velocity clearly changes if we consider a
walker with initial position $Y_i < 0$.  Hence the correct limiting
form of the function $F_i(\tau_0/\tau^\prime, \tau^\prime/\tau)$ in
equation (\ref{guess1}) is
\begin{equation}
F_i(0,0) =  \frac{2 \sigma_i}{\sqrt{\pi}}
\end{equation}
in which $\sigma_i = \sgn(Y_i)$.

The two-point velocity correlation function $\avgR{\dot{X}_i(\tau_1)
\dot{X}_i(\tau_2)}$ is obtained in a similar way.  We defer the
details of the calculation to the appendix, quoting here only the
result over the required range $\tau_0 \ll \tau_1, \tau_2 \ll \tau$
which is
\begin{equation}
\label{correlR}
\avgR{\dot{X}_i(\tau_1) \dot{X}_i(\tau_2)} = 2 \left[ \delta(\tau_2 - \tau_1)
+ \frac{2}{\pi \tau_+} \sqrt{\frac{\tau_+-\tau_-}{\tau_-}} \right] \;.
\end{equation}
Recall that $\tau_- = \min\{\tau_1,\tau_2\}$ and $\tau_+ =
\max\{\tau_1,\tau_2\}$.  Note that this expression is in agreement
with the proposal (\ref{guess2}), and reveals the limiting form of the
function $G$ to be
\begin{equation}
G_i(0,\tau_-/\tau_+,0) = \frac{2}{\pi} \sqrt{\frac{\tau_+}{\tau_-}-1} \;.
\end{equation}
Once again, the dependence of this quantity on the walker's initial
position $Y_i$ has dropped out; furthermore, the two-point correlation
function is independent of the sign of $Y_i$, so we can drop the
subscript $i$ on $G$.

We now consider the first order term in (\ref{thetaexpand}).  We have
\begin{eqnarray}
\avgR{S_1}-\avgU{S_1} &=& \frac{1}{4} \psum_{i,j} \int_0^\tau
\Di{\tau^\prime} \avgR{\dot{X}_i(\tau^\prime)}
\avgR{\dot{X}_j(\tau^\prime)} \nonumber\\
&=& \frac{1}{4} \psum_{i,j} \int_0^\tau
\frac{\D{\tau^\prime}}{\tau^\prime} F_i(\tau_0/\tau^\prime,
\tau^\prime/t) F_j(\tau_0/\tau^\prime, \tau^\prime/t) \;.
\end{eqnarray}
The unrestricted averages have disappeared from the right-hand side of
this expression because $\avgU{\dot{X}_i} = 0$.  Now, using equation
(\ref{limit1}), we find that the limiting form of this expression
required in the expansion (\ref{thetaexpand}) reads
\begin{equation}
\fl \lim_{\tau\to\infty} \tau \frac{\D}{\D\tau} \left(
\avgR{S_1}-\avgU{S_1} \right) = \frac{1}{4} \psum_{i,j} F_i(0,0)
F_j(0,0) = \frac{1}{\pi} \psum_{i,j} \sigma_i \sigma_j \;.
\end{equation}
To perform the remaining summation, we note that $\sigma_i \sigma_j$
is equal to $+1$ if walkers $i,j$ both start on the same side of the
origin, and is equal to $-1$ if they start on opposite sides of the
origin.  If there are $N_L$ ($N_R$) walkers initially to the left
(right), there are $N_L(N_L-1) + N_R(N_R-1)$ positive terms in the
summation and $2 N_L N_R$ negative terms.  Thus one finds that the
coefficient of $\alpha_N$ in (\ref{thetaexpand}) is
\begin{equation}
\lim_{\tau\to\infty} \tau \frac{\D}{\D\tau} \left(\avgR{S_1} -
\avgU{S_1}\right) = \frac{1}{\pi} (\Delta^2-N)
\end{equation}
in which $\Delta = N_L - N_R$.

Let us now turn to the second-order term in (\ref{thetaexpand}).  From
(\ref{variance}) we have the general form
\begin{eqnarray}
\fl \frac{1}{2} \left( \avgR{S_1^2} - \avgR{S_1}^2 - \avgU{S_1^2} +
\avgU{S_1}^2 \right) = \nonumber\\
\fl\quad \frac{1}{32} \int_0^\tau \Di{\tau_1} \int_0^\tau
\Di{\tau_2} \left[ 4 \psum_{i,j,k} A_{ijk}(\tau_0/\tau_1,\tau_1,\tau_2,\tau_2/\tau) + 2 \psum_{i,j}
B_{ij}(\tau_0/\tau_1,\tau_1,\tau_2,\tau_2/\tau) \right] \;.
\end{eqnarray}
In the regime $\tau_0 \ll \tau_1, \tau_2 \ll t$ we have from (\ref{meanvarU}),
(\ref{meanR}) and (\ref{correlR})
\begin{eqnarray}
\fl A_{ijk}(0,\tau_-,\tau_+,0) = \frac{8 \sigma_j \sigma_k}{\pi^2}
\left[ \frac{\pi \delta(\tau_2-\tau_1)}{\tau_2} + \frac{2}{\tau_+^2}
\frac{\tau_+}{\tau_-} \left( \sqrt{1-\frac{\tau_-}{\tau_+}} -1 \right) \right] \\
\fl B_{ij}(0,\tau_-,\tau_+,0) = - \frac{16}{\pi^2 \tau_+^2}
\;.
\end{eqnarray}
As with the first order term, we calculate the limit in
(\ref{thetaexpand}) by making use of the results (\ref{limit1}) and
(\ref{limit2}).  This procedure yields the result
\begin{eqnarray}
\fl \lim_{\tau\to\infty} \tau \frac{\D}{\D{\tau}} \frac{1}{2} \left(
\avgR{S_1^2} - \avgR{S_1}^2 - \avgU{S_1^2} + \avgU{S_1}^2 \right) =
\nonumber\\
\frac{1}{\pi^2} \left[ \left( \pi + 4 \int_0^1 \Di{u}
\frac{\sqrt{1-u}-1}{u} \right) \psum_{i,j,k} \sigma_j \sigma_k - 2
\psum_{i,j} \int_0^1 \Di{u} \right] \;.
\end{eqnarray}
To obtain the final expression for the expansion (\ref{thetaexpand})
we use the fact that
\begin{equation}
\int_0^1 \Di{u} \frac{\sqrt{1-u}-1}{u} = 2 (\ln 2 - 1) \;.
\end{equation}
Also, we must consider the possible ways of choosing three particles
labelled $i,j,k$ from a set of $N$ particles comprising a number $N_L$
with $\sigma=-1$ and $N_R$ with $\sigma=1$.  These considerations lead
us to
\begin{equation}
\psum_{i,j,k} \sigma_j \sigma_k = (N-2) (\Delta^2 - N) \;.
\end{equation}
Similarly the number of terms in the sum $\sum^\prime_{i,j}$ is just
$N(N-1)$ which finally allows us to write down an expression for the
exponent $\theta$ which constitutes the main result of this paper.  It
reads
\begin{eqnarray}
\label{thetaresult}
\fl \theta = \frac{N}{2} + \frac{1}{\pi} (N - \Delta^2) \alpha_N \nonumber\\
+ \frac{1}{\pi^2} \left[ (N-2) (N-\Delta^2) (\pi - 8(1-\ln 2)) + 2 N
(N-1) \right] \alpha_N^2 \;.
\end{eqnarray}

As a check of this formula, we consider the exactly solvable cases of
$N=1$ and $N=2$.  When $N=1$, one can simply transform to a frame in
which the particle is stationary.  Then, its survival probability is
given by equation (\ref{Q1}) with $D$ replaced by $D+D^\prime$ which
implies a persistence exponent of $\theta=\frac{1}{2}$.  Since
$\Delta^2 = 1$ when $N=1$, we find that the perturbative expression
for $\theta$ (\ref{thetaresult}) also gives the exact result
$\theta=\frac{1}{2}$.

When $N=2$, there are two possible arrangements of the walkers.
Either they surround the particle, in which case $\Delta=0$ and the
exact exponent is given by equation (\ref{exactsurround}), or they are
both on one side of the particle, which has $\Delta=\pm 2$ and
$\theta$ is given by equation (\ref{exactchase}).  Expanding these
exact expressions as series in $\alpha_2$ one finds
\begin{equation}
\theta = 1 \pm \frac{2}{\pi} \alpha_2 +
\frac{4}{\pi^2} \alpha_2^2 + \cdots
\end{equation}
where the plus sign is taken for the case $\Delta=0$ and the minus
sign for $\Delta=\pm 2$.  Thus to second order in $\alpha_2$ we find
agreement between the perturbation series (\ref{thetaresult}) and the
exactly known results.

\section{Discussion and conclusion}
\label{secconc}

In this work we have studied the survival probability of a diffusing
particle in the presence of a finite number of mobile traps.  In the
absence of a complete exact solution for the problem, we have devised
a method for calculating the persistence exponent $\theta$ [defined by
equation (\ref{thetadef})] valid when the particle's diffusion
constant is small.  In order to calculate the coefficients in the
expansion, it is necessary to find velocity correlation functions for
a random walker in the presence of an absorbing boundary.  In this
paper, we calculated the exponent $\theta$ to second order,
culminating in the expression (\ref{thetaresult}).  We stress that, in
principle, one could go to higher order, although the amount of work
required to obtain $n$-time velocity correlation functions is likely
to increase rapidly with $n$.

We wish to make a few remarks about our result (\ref{thetaresult}).
First, as we have already noted, the expansion parameter $\alpha_N$
defined by equation (\ref{alphadef}) has an interpretation in the
context of the (hyper)wedge geometry that has been previously used to
obtain results for this problem.  Although the connection appears at
first sight intriguing, it is probably no coincidence since in the two
approaches the parameter arises through a rescaling that gives rise to
uncorrelated diffusion of the traps with diffusion constant unity.
However we do see that a system with perturbed equations of motion
(treated using the path-integral approach) can be related to one with
the original equations of motion but perturbed boundaries (i.e.\ the
hyperwedge geometry).

Secondly, we find that the expansion to second order
(\ref{thetaresult}) depends on two parameters: the total number of
traps in the system $N$ and the square difference $\Delta^2$ in the
number of traps initially placed on either side of the particle.  The
origin of this observation lies in the fact that at sufficiently large
times the initial position of a particular trap is unimportant.  Of
course, the number of traps initially positioned to each side of the
particle plays an important role, and enters into the perturbation
expansion through summations in which some terms are positive and some
negative depending on the initial arrangement of the traps.  We would
expect, therefore, the combination $N-\Delta^2$ to appear in
higher-order terms in the expansion.

We also note from the expansion (\ref{thetaresult}) that when the
asymmetry is small, i.e.\ $\Delta^2 < N$, the exponent $\theta$ is an
increasing function of the particle's diffusion constant $D^\prime$ at
fixed trap diffusion constant $D$.  Hence, in this near-symmetric
situation, the particle is more likely to survive longer if it is at
rest than if it is slowly diffusing.  Of course, we do not know if
$\theta$ continues to increase for large $D^\prime$ since we only have
the first two terms in the perturbation expansion.  However, our
intuition suggests that this be the case.

The case of large asymmetry, $\Delta^2 > N$ is more subtle since one
of the terms in the coefficient of $\alpha_N^2$ is always positive.
We learn a little by defining the critical asymmetry $\Delta_c$ at
fixed $N$ and $\alpha_N$ to be that for which the persistence exponent
$\theta = \frac{N}{2}$.  Then for $\Delta>\Delta_c$ one has a regime
in which the particle's survival probability is not (locally)
maximised by remaining still.  One can understand this phenomenon from
the extreme case of all traps being on one side of the particle
($\Delta=\pm N$).  Then, there is some probability for the particle to
diffuse into the region which is devoid of traps, thereby increasing
its survival probability.  By rearranging the expansion
(\ref{thetaresult}) one finds for small $\alpha_N$ that
\begin{equation}
\Delta_c^2 = N + \frac{2N(N-1)}{\pi} \alpha_N + \mathrm{O}(\alpha_N^2) \;.
\end{equation}
Of course, it is again difficult to make any statements for
large $\alpha_N$ given only the first two terms of the perturbation
series.

The final comment we wish to make about the form of
(\ref{thetaresult}) is that, although the quantity $\alpha_N$ seems to
be the natural expansion parameter in the problem, the presence of a
nontrivial coefficient in the second order term is not suggestive of
there being a simple, closed expression valid for all values of the
parameters $N$, $\Delta$, $D$ and $D^\prime$.  Nevertheless, one might
find that the marginal case $N=\Delta^2$ is in some way much easier to
understand.

The trapping reaction studied in this work suggests a couple of
generalisations.  Firstly there is the question of how to treat higher
dimensions.  In a forthcoming work \cite{BB02c}, we extend the methods
introduced in \cite{BB02a} for the case of an infinite sea of traps to
dimensions greater than one.  It would be interesting to see if the
perturbation theory could be applied to higher dimensions to treat the
case of a finite number of traps.  Finally, we have studied in this
work the trapping reaction on an infinite line.  It appears that
the problem in the periodic system is much harder to treat than the
case considered here.

\ack

RAB acknowledges financial support from the EPSRC under grant
GR/R53197.

\appendix

\section{Two-point velocity function for surviving walks}

In this appendix we outline the derivation of the correlation function
$\avgR{\dot{X}(\tau_1) \dot{X}(\tau_2)}$ as given by equation
(\ref{correlR}).  Recall that this is the two-point velocity
correlation function for a single random walker that has a diffusion
constant $D_0$ and does not cross the origin at any time $0 \le
\tau^\prime \le \tau$, where $\tau > \tau_1, \tau_2$.  We shall
assume, with no loss of generality, that the walker's initial position
$Y > 0$ and that $\tau_1<\tau_2$.

The starting point is to consider the probability $P_1(\tau; X_1, \tau_1;
X_2, \tau_2 | Y)$ for the walker to be at $X_1$ at time $\tau_1$, $X_2$ at
time $\tau_2$ and to survive until time $t$ given that it started at $Y$.
Then,
\begin{eqnarray}
\label{X1X2}
\fl\avgR{X(\tau_1) X(\tau_2)} = \int_0^\infty \Di{X_1} \int_0^\infty
\Di{X_2} X_1 X_2 \, \frac{P_1(\tau; X_1, \tau_1; X_2, \tau_2 |
Y)}{Q_1(\tau | Y)} \nonumber\\
\fl\quad = \int_0^\infty \Di{X_1} \int_0^\infty \Di{X_2} X_1 X_2 \,
\frac{Q_1(\tau-\tau_2|X_2) G_1(X_2, \tau_2-\tau_1 | X_1) G_1(X_1, \tau_1 | Y)}
{Q_1(\tau | Y)} \;,
\end{eqnarray}
in which the Markovian property of the diffusion process has been used
to write the joint probability distribution function $P_1(\tau; X_1, \tau_1;
X_2, \tau_2 | Y)$ in terms of the functions $G_1(X, \tau | Y)$ and $Q_1(\tau |
Y)$ defined through equations (\ref{G1}) and (\ref{Q1}).  Note that it
was also necessary to normalise the distribution according to the
probability that a walker starting at $Y$ survives a time $\tau$.

Assuming that $\tau_0 = Y^2 \ll \tau_1, \tau_2 \ll \tau$, we can
expand both the numerator and denominator in (\ref{X1X2}) to first
order in $Y/\sqrt{\tau_1}$ and $X_2/\sqrt{\tau}$ to find
\begin{eqnarray}
\fl \avgR{X(\tau_1) X(\tau_2)} = \frac{1}{4\pi \tau_1 \sqrt{\tau_1
(\tau_2-\tau_1)}} \int_0^\infty \Di{X_1} \int_0^\infty \Di{X_2} X_1^2
X_2^2 \exp\left( - \frac{X_1^2}{4 \tau_1} \right) \nonumber\\
\lo\times \left[ \exp\left( -
\frac{(X_2-X_1)^2}{4 (\tau_2-\tau_1)} \right) - \exp\left( -
\frac{(X_2+X_1)^2}{4 (\tau_2-\tau_1)} \right) \right] \;.
\end{eqnarray}
This integral can be evaluated without recourse to further
approximations.  First one integrates over $X_2$, which yields
\begin{eqnarray}
\fl \avgR{X(\tau_1) X(\tau_2)} = \frac{1}{2 \sqrt{\pi} \tau_1^{3/2}}
\int_0^\infty \Di{X_1} X_1^2 \exp\left(-\frac{X_1^2}{4\tau_1}\right)
\bigg[ (X_1^2 + 2 (\tau_2-\tau_1)) 
\nonumber\\
\lo\times  \erfc\left( \frac{X_1}{\sqrt{4
(\tau_2-\tau_1)}} \right) + \sqrt{\frac{4(\tau_2-\tau_1)}{\pi}} X_1
\exp \left( -\frac{X_1^2}{4(\tau_2-\tau_1)} \right) \bigg] \;.
\end{eqnarray}

To perform the remaining integration, it is helpful to make a change
of variable $X_1 = 2 \sqrt{\tau_2-\tau_1} u$ which leads to
\begin{equation}
\fl \avgR{X(\tau_1) X(\tau_2)} = \frac{16 (\tau_2-\tau_1)^{5/2}}{\sqrt\pi
\tau_1^{3/2}} \int_0^\infty \Di{u} \left[ \frac{1}{\sqrt\pi} u^3
\E^{-\alpha u^2} + \frac{1}{2} u^2 \E^{-\beta u^2} \erf u + u^4
\E^{-\beta u^2} \right]
\end{equation}
in which $\alpha = \tau_2/\tau_1$ and $\beta =
(\tau_2-\tau_1)/\tau_1$.  Then, using
\begin{eqnarray}
\fl
\int_0^\infty \Di{u} u^3 \E^{-\alpha u^2} = \frac{1}{2\alpha^2} 
\\
\fl
\int_0^\infty \Di{u} u^{2n} \E^{-\beta u^2} \erf u = \left( -
\frac{\D}{\D\beta} \right)^n \int_0^\infty \Di{u} \E^{-\beta u^2} \erf
u = \left( - \frac{\D}{\D\beta} \right)^n
\frac{\arctan(1/\sqrt\beta)}{\sqrt{\pi\beta}}
\end{eqnarray}
we obtain, after a lot of manipulation,
\begin{equation}
\fl \avgR{X(\tau_1) X(\tau_2)} = \frac{4\tau_1}{\pi}
\left[ \left( 2+ \frac{\tau_2}{\tau_1} \right) \arctan
\sqrt{\frac{\tau_1}{\tau_2-\tau_1}} + 3 \sqrt{\frac{\tau_2-\tau_1}{\tau_1}} \right] \;.
\end{equation}

To find the velocity correlation function $\avgR{\dot{X}(\tau_1)
\dot{X}(\tau_2)}$ we differentiate with respect to $\tau_1$ and $\tau_2$.  We
find
\begin{equation}
\label{dX1X2dt2}
\fl
\frac{\partial}{\partial \tau_2} \avgR{X(\tau_1) X(\tau_2)} = \frac{4}{\pi}
\left\{ \begin{array}{ll}
\arctan \sqrt{\frac{\tau_1}{\tau_2-\tau_1}} + \frac{\sqrt{\tau_1(\tau_2
-\tau_1)}}{\tau_2} & \tau_1 < \tau_2 \\
2 \arctan \sqrt{\frac{\tau_2}{\tau_1-\tau_2}} + 2\sqrt{\frac{\tau_1-\tau_2}{\tau_2}} &
\tau_1 > \tau_2 \end{array}  \;. \right.
\end{equation}
For $\tau_1<\tau_2$ one has
\begin{equation}
\avgR{\dot{X}(\tau_1)\dot{X}(\tau_2)} = \frac{4}{\pi \tau_2}
\sqrt{\frac{\tau_2-\tau_1}{\tau_1}}
\end{equation}
and the same expression with $\tau_1 \leftrightarrow \tau_2$ when $\tau_1>\tau_2$.
At $\tau_1 = \tau_2$, however, there is a jump in (\ref{dX1X2dt2}) of height
$2$ which implies that
\begin{equation}
\avgR{\dot{X}(\tau_1)\dot{X}(\tau_2)} = 2 \left[ \delta(\tau_2-\tau_1)
+ \frac{2}{\pi \tau_+} \sqrt{\frac{\tau_+-\tau_-}{\tau_-}} \right] \;.
\end{equation}
Thus we conclude our derivation of (\ref{correlR}).

%
%

\end{document}